\documentclass[12pt,preprint]{aastex}

\shorttitle{$K_s$ band luminosity evolution of the AGB population}
\shortauthors{Ko et al.}

\begin{document}


\title{$K_s$ BAND LUMINOSITY EVOLUTION OF THE ASYMPTOTIC GIANT BRANCH  POPULATION BASED ON STAR CLUSTERS IN THE LARGE MAGELLANIC CLOUD}


\author{Youkyung Ko,  Myung Gyoon Lee, and Sungsoon Lim}
\affil{Astronomy Program, Department of Physics and Astronomy, Seoul National University,
Seoul 151-742, Korea}

\email{ykko@astro.snu.ac.kr, mglee@astro.snu.ac.kr, slim@astro.snu.ac.kr}


\begin{abstract}

We present a study of $K_s$ band luminosity evolution of the asymptotic giant branch (AGB) population in simple stellar systems using star clusters in the Large Magellanic Cloud (LMC). We determine physical parameters of LMC star clusters including center coordinates, radii, and foreground reddenings. Ages of 83 star clusters are derived from isochrone fitting with the Padova models, and those of 19 star clusters are taken from the literature. The AGB stars in 102 star clusters with log(age) = 7.3 -- 9.5 are selected using near-infrared color magnitude diagrams based on 2MASS photometry. Then we obtain the $K_s$ band luminosity fraction of AGB stars in these star clusters as a function of ages. 
The $K_s$ band luminosity fraction of AGB stars increases, on average, as age increases from log(age) $\sim$ 8.0, reaching a maximum at log(age) $\sim$ 8.5, and it decreases thereafter. There is a large scatter in the AGB luminosity fraction for given ages, which is mainly due to stochastic effects. 
We discuss this result in comparison with five simple stellar population models.
The maximum $K_s$ band AGB luminosity fraction for bright clusters is reproduced by the models that expect the value of 0.7 -- 0.8 at log(age) = 8.5 -- 8.7.
We discuss the implication of our results with regard to the study of size and mass evolution of galaxies.

\end{abstract}





\section{Introduction}

The asymptotic giant branch (AGB), a representative intermediate-age population, is the most luminous evolutionary stage for low- and intermediate-mass (1 -- 8 M$_{\odot}$) stars. AGB stars emit a huge amount of near- to mid-infrared light because of their low effective temperature ($\mathrm{T_{eff}} \sim$ 1000 -- 4000 K) and circumstellar dust. Hence it is expected that they contribute significantly to the infrared light of galaxies. However, their infrared luminosity contribution in galaxies, even in a simple stellar population (SSP), is still a debating issue with regard to constructing evolutionary population synthesis (EPS) models (\citealt{mar11}, \citealt{bru13}, \citealt{noe13}, and references therein).

The EPS models are an useful tool to determine physical parameters of stellar complexes such as star clusters and galaxies by spectral energy distribution (SED) fitting. These models reproduce SEDs of stellar systems by synthesizing the SEDs of stellar populations in various evolutionary stages, assuming age, metallicity, foreground and internal reddening values, star formation history, and so on. 
Therefore, the spectral energy contribution of individual stellar populations influences  the estimation of physical parameters of stellar systems, and especially that of the AGB population plays a key for understanding of the SEDs for stellar systems expected by EPS models.

There are various EPS models that include the AGB evolutionary process (e.g., \citealt{cb91}, \citealt{bc93}, \citealt{fr97}, \citealt{mar98}, \citealt{lei99}, \citealt{vaz99}, \citealt{sch02}, \citealt{bc03}, \citealt{jim04}, \citealt{mar05}, \citealt{con09}, \citealt{kot09}, \citealt{cg10}, \citealt{vaz10}, \citealt{ms11}).
Bruzual \& Charlot (2003, hereafter BC03), for example, presented an EPS model based on Padova evolutionary tracks including just two evolutionary stages for thermally pulsing AGB (TP-AGB) \citep{gir00}, while their recent model includes 15 evolutionary stages of the TP-AGB oxygen-rich and carbon-rich phases \citep{bru10}.
The Padova evolutionary tracks adopted in the BC03 model include a given core overshooting efficiency. On the other hand, Maraston (1998, 2005) used evolutionary tracks assuming zero efficiency of the core overshooting from Cassisi et al.(1997, 2000) to determine luminosity contribution of main-sequence populations. It makes the lifetime of main-sequence stars in the \citet{mar05} model shorter compared with Padova isochrones. In addition, she used the fuel consumption theorem \citep{rb86} for post-main-sequence stars, not adopting the isochrone synthesis method.
In the model of \citet{mar98}, the TP-AGB evolutionary process is modified in the fuel consumption theorem, applying the advanced envelope burning process of AGB stars. \citet{rb86} estimated the contribution of TP-AGB population in the young SSPs with ages $< 10^8$ yr, and compared it with the star clusters in the the Magellanic Clouds (MCs). They found that it is too high to be consistent with the observational results based on these star clusters. 
Considering this difference, \citet{mar05} suggested a revised model, assuming shorter lifetime and lower fuel consumption of TP-AGB stars than \citet{rb86}.

BC03 and \citet{mar05} adopted different stellar population synthetic methods, and the SEDs reproduced by these models show recognizable differences. A number of studies argue whether or not these EPS models reproduce well observational quantities of stellar systems related with the AGB population as follows. 
Some studies found that spectra and SEDs of post-starburst galaxies or distant galaxies (z $\sim$ 1 -- 2) have no near-infrared boosted features, which is similar to the expectation of the BC03 model (\citealt{muz09}, \citealt{kri10}, \citealt{zib13}). 
In addition, \citet{cg10} suggested that the BC03 model and their own EPS model, a flexible stellar population synthesis model, reproduce the color of star clusters and post-starburst galaxies better than the \citet{mar05} model. 
In contrast, others confirmed that the \citet{mar05} model performs well in reproducing near-infrared colors, $(Y-K)$ and $(H-K)$, and SEDs of low to high redshift galaxies (\citealt{mar06}, \citealt{emi08}, \citealt{mac10}). Moreover, \citet{van06} and \citet{hen11} showed that the rest-frame $K$ band mass-to-light ratio evolution and $K_s$ band luminosity evolution of galaxies are explained better by the \citet{mar05} model than by the BC03 model.

In addition to these galaxy studies, there are several studies investigating star clusters, especially star clusters in the MCs, 
considering the relevance between the model performance and ages of stellar systems. 
Because of their proximity, they are very useful to examine not only the individual stellar population directly but also the integrated properties of star clusters. Therefore, they are an ideal object to calibrate SSP models.
\citet{pes08} investigated optical to near-infrared colors, $(B-J)$, $(V-J)$, and $(J-K)$, of 54 star clusters in the MCs, and concluded that the BC03 and the \citet{mar05} present the best performance for intermediate-age (0.2 -- 2 Gyr) and old ($>$ 2 Gyr) star clusters, respectively. 
Lyubenova et al. (2010, 2012) analyzed integrated near-infrared high-resolution spectra of six globular clusters in the Large Magellanic Cloud (LMC) with ages $\sim$ 1 -- 13 Gyr. They showed that the \citet{mar05} model expects $J$ and $H$ band spectra of all their sample star clusters adequately, while it does not for $K$ band spectra of the star clusters with ages $<$ 2 Gyr.
Recently \citet{noe13} presented the calibration data for stellar population models using 43 star clusters in the MCs. They compared the observed colors of the star clusters, $(V-K)_{0}, (J-K)_{0}$, and $(V-I)_{0}$, as a function of ages with the theoretical expectation of various models including BC03 and \citet{mar05}. In conclusion, for the ages older than 1 Gyr, the models of \citet{mar05} and BC03 are along the upper end lower end of the observed color, respectively. In the case of the younger ages, BC03 model reproduces well the observed color of star clusters, while \citet{mar05} model expects them too red.

While above studies investigated the integrated properties of star clusters, \citet{muc06} analyzed the resolved AGB population in the LMC star clusters in detail.
They investigated 19 LMC star clusters in terms of $K_s$ band luminosity contribution of AGB stars as a function of cluster ages. The star clusters used in \citet{muc06} have ages of 100 Myr -- 3 Gyr (log(age) $\sim$ 8.0 -- 9.5). They concluded that their empirical results are consistent with the expectation of \citet{mar05} for this age range. Later, \citet{muc09} suggested a similar conclusion from the additional study of four old star clusters in the Small Magellanic Cloud (SMC). 

In this study, we investigate the evolution of AGB luminosity contribution with a large number of LMC star clusters with log(age) $\sim$ 7.3 -- 9.5, overcoming the shortcomings in the previous studies.
Because Mucciarelli et al. (2006, 2009) included few star clusters with log(age) $\sim$ 8.3 -- 8.7 (see Figure 5 in \citealt{muc09}), we enlarged the number of star cluster samples with this age range. In this age range, the AGB luminosity contribution is expected to change rapidly according to the \citet{mar05} model.
In addition, most of previous star cluster studies adopted star cluster ages from various literature. We determine ages of LMC star clusters by isochrone fitting homogeneously, using the resolved stars.

This paper is organized as follows. In \S2, we introduce the photometric data and images used in this study. 
\S3 describes the method to estimate physical parameters of sample star clusters such as center coordinates, radii, ages, and foreground reddenings including the cluster sample section.
In \S4, we select AGB stars in each star cluster and derive the $K_s$ band luminosity contribution of AGB stars to total luminosity of star clusters as a function of ages. 
In \S5, we compare our results with previous studies, and also compare the primary results with the theoretical expectation from EPS models, including the discussion of stochastic effects.
Final section summarizes the main results and presents the conclusion of this study.

\section{Data}

\citet{bic08} presented an extended source catalog of the MCs including star clusters, emission nebulae, associations, and HI shells. They compiled data from various literature including the findings based on photographic survey plates. It contains center coordinates (R.A. and Declination), major and minor axes, and position angles of 3,700 star clusters of the MCs including the LMC, SMC, and the Magellanic Bridge regions. 
We redetermined  the centers and radii of LMC star clusters based on the center coordinates presented in \citet{bic08}.

We used optical ($UBVI$) and near-infrared ($JHK_s$) point source catalogs of the LMC. \citet{zar04} presented $U, B, V$, and $I$ band photometry of 24,107,004 point sources in the central 64 deg$^2$ of the LMC from the Magellanic Clouds Photometric Survey (MCPS; \citealt{zar97}). The MCPS obtained drift-scan images using 1-m Las Campanas Swope Telescope and Great Circle Camera \citep{zar96}. Their photometry is incomplete below 21.5, 23.5, 23, and 22 mag in $U, B, V$, and $I$ bands in sparse regions, respectively. This catalog is used to determine center coordinates, radii, ages, and foreground reddenings of the star clusters.

In order to distinguish AGB stars from other populations, we used a near-infrared point source catalog from the two micron all sky survey (2MASS; \citealt{skr06}). 
In the central 100 deg$^2$ region of the LMC (10$\arcdeg \times$ 10$\arcdeg$), there are 1,430,676 point sources detected in the 2MASS. 
The limiting magnitudes of photometry are 15.8, 15.1, and 14.3 mag in $J, H$, and $K_s$ bands, respectively, which correspond to 10$\sigma$ point source detection level. The AGB stars in the LMC are brighter than $K_s \sim 12.3$ mag, much brighter than the limiting magnitudes.
We also used 2MASS Atlas Images to estimate the integrated luminosity of the star clusters,
retrieving the $K_s$ band images of our sample clusters using 2MASS interactive image service\footnote{http://irsa.ipac.caltech.edu/applications/2MASS/IM/interactive.html}.

\section{Physical Parameters of LMC Star Clusters and Sample Selection}

	\subsection{Center Coordinates and Sizes of Star Clusters}

\begin{figure}[hbt]
\epsscale{1}
\plotone{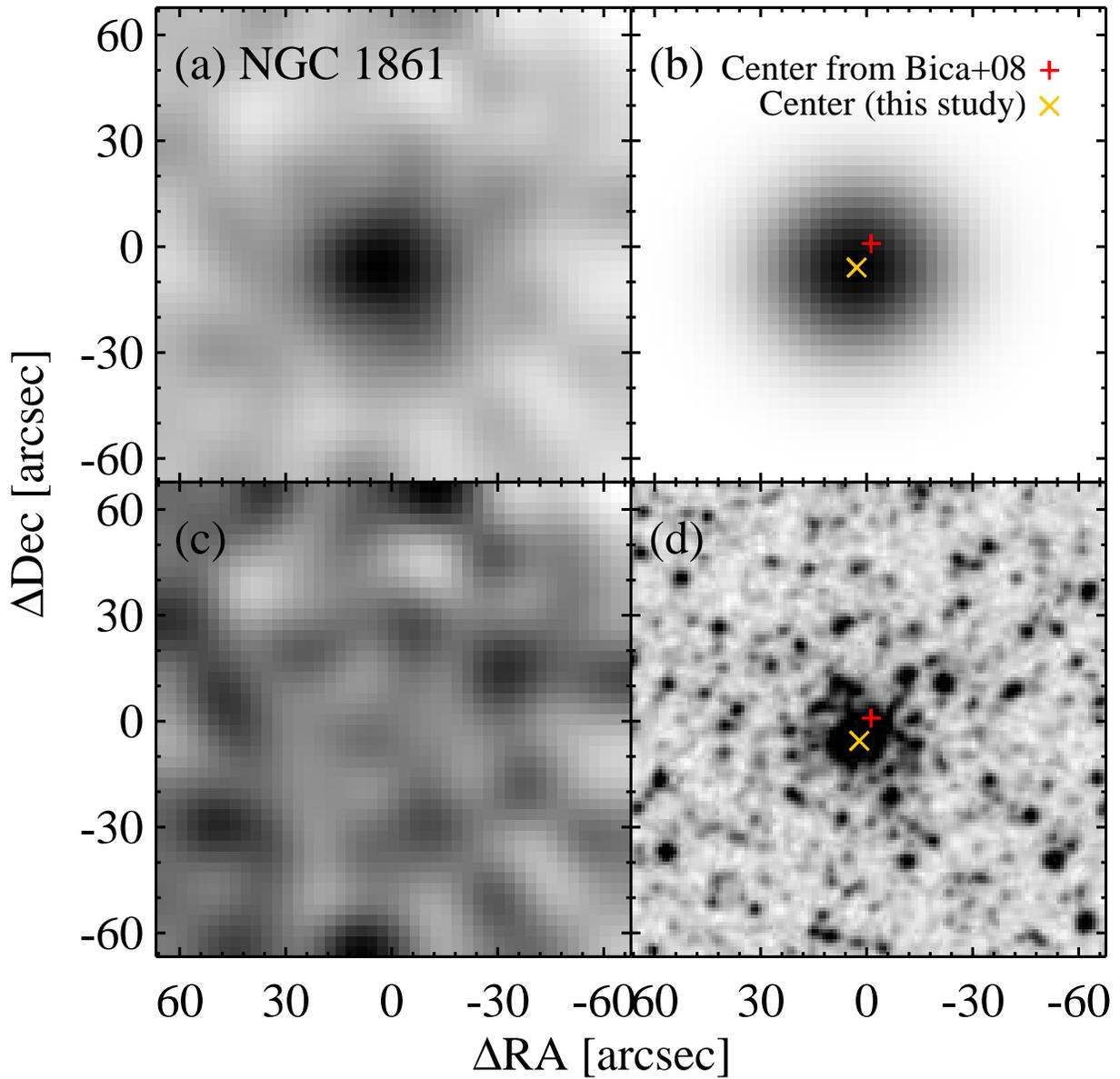}
\caption{An example for determination of the center of a star cluster, NGC 1861. (a) 2-dimensional number density map smoothed with a boxcar filter for the bright stars with $V < 20.5$ mag around NGC 1861 derived from the MCPS catalog of \citet{zar04},
 (b) Fitted number density map with 2-dimensional Gaussian fitting. The plus and cross mark, respectively, represent the cluster center presented by \citet{bic08} and that derived in this study. (c) Residual number density map derived from subtraction of (b) from (a).  (d) A gray scale map of a digitized sky survey $R$ band image. The plus and cross mark represent the same as in (b).  \label{fig:center}}
	\end{figure}

We determined centers for 1,645 star clusters and radii for 1,708 star clusters, respectively, using the MCPS catalog \citep{zar04} as a part of our study of LMC star clusters. We constructed 2-dimensional number density maps of bright stars with $V < $ 20.5 mag around the center coordinates of the star clusters presented by \citet{bic08}. The number density maps are smoothed with a boxcar filter whose width is 20$\arcsec$. Center coordinates, coordinate errors, and position angles of each star cluster were estimated by 2-dimensional Gaussian fitting of this smoothed number density map. The field of view for the fitting region is 3 times larger than the radius of each cluster given in \citet{bic08}.
Figure \ref{fig:center} shows an example of centering process for one cluster NGC 1861. 
We attempted 2-dimensional Gaussian fitting to 3,064 LMC star clusters in \citet{bic08}, but stellar number density maps could obtain acceptable fits for only 1,645 star clusters. The fitting results are not reliable in the case of poor, faint, or binary clusters, in which case that the center coordinates presented by \citet{bic08} are adopted.

\begin{figure}[hbt]
\epsscale{1}
\plotone{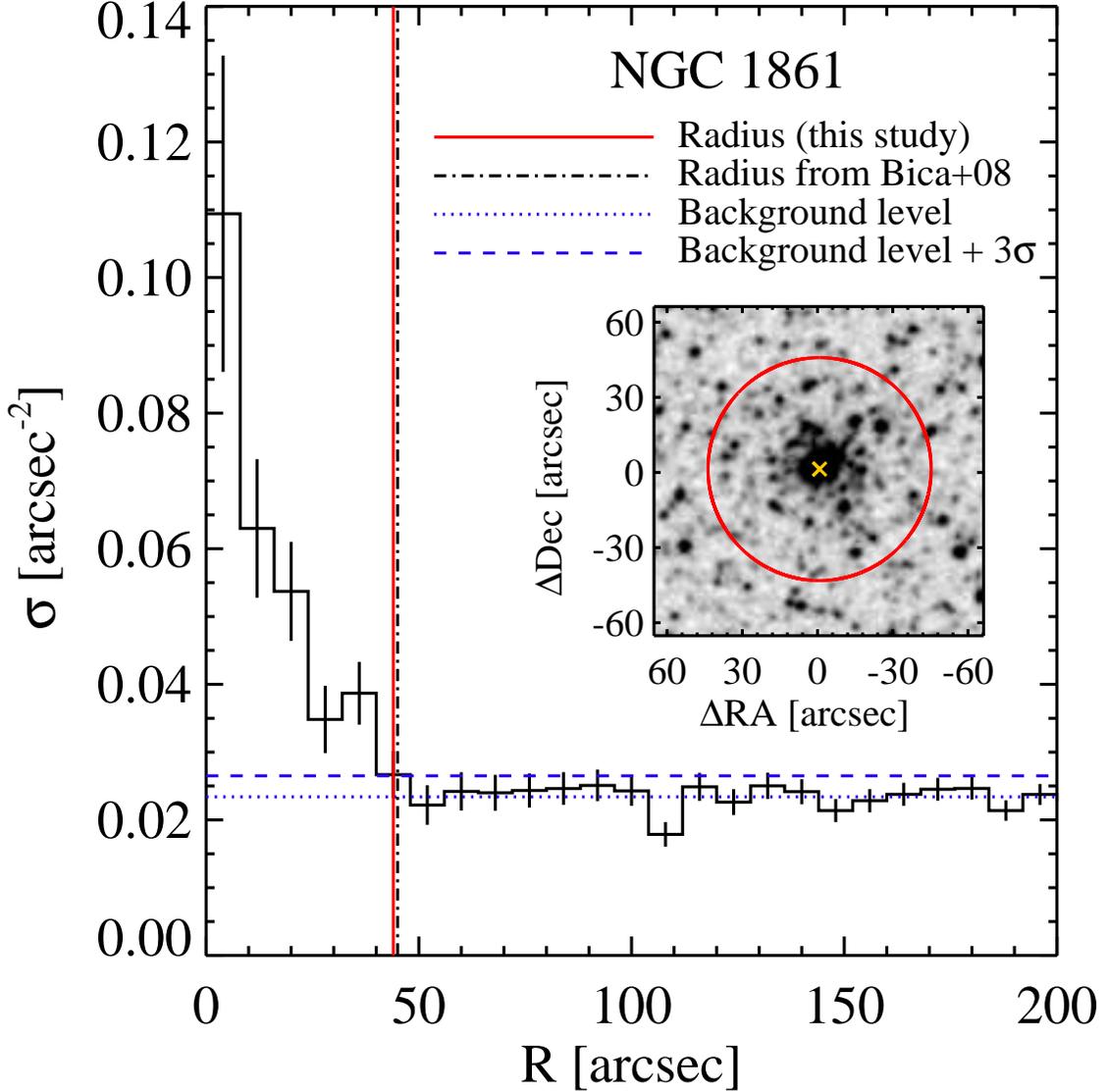}
\caption{Radial number density profile for the bright stars with $V <20.5$ mag around NGC 1861 in the MCPS catalog of \citet{zar04}. Error bars indicate the Poisson errors estimated in each bin. 
The solid line and  dot-dashed line represent the cluster radius presented by \citet{bic08} and that determined in this study, respectively. The background level (dotted line) is the median value of the number density of the stars located in the annulus, 200$\arcsec <$ R $<$ 300$\arcsec$.  We adopted 3$\sigma$ upper level (dashed line) to determine the radius of each cluster (see texts). The stamp image is a gray scale map of a digitized sky survey $R$ band image of NGC 1861, showing the cluster size determined in this study by a large circle.\label{fig:radius}}
	\end{figure}

The radius of each cluster was determined from radial number density profiles (see Figure \ref{fig:radius}). The radial number density profile is obtained by counting point sources with $V <$ 20.5 mag. We estimated a median value of the background number density ($\mathrm{n_{bg}}$) of stars located between 200$\arcsec$ and 300$\arcsec$ from the center of each cluster. The standard deviation from $\mathrm{n_{bg}}$ ($\mathrm{\sigma_{bg}}$) is calculated, and the area of which number density greater than $\mathrm{n_{bg} + 3\sigma_{bg}}$ is considered as a cluster area. Finally, we determined radii of 1,708 LMC star clusters, and radii of the other clusters that do not show a prominent concentration are adopted from \citet{bic08}. Most of these clusters are poor, faint, or binary ones. \citet{bic08} presented major and minor axes of star clusters, from which we define the radius of each star cluster as the mean value of semi-major and semi-minor axes. 

\begin{figure}[hbt]
\epsscale{1}
\plotone{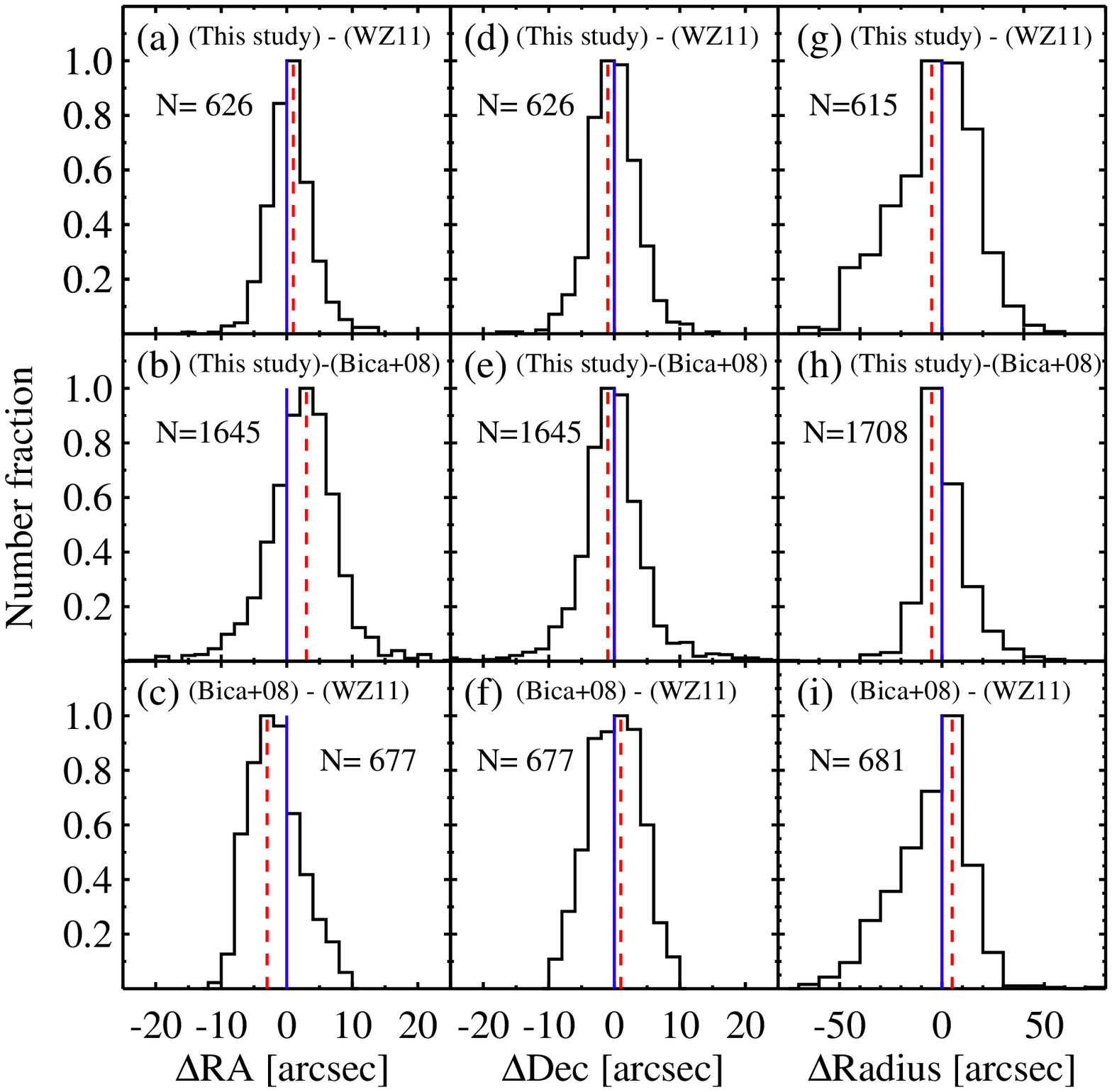}
\caption{Comparison of center coordinates and radii of star clusters between this study and other references, \citet{wz11} (a,d, and g panels) and \citet{bic08} (b, e, and h panels). The bottom three panels show the differences between \citet{bic08} and \citet{wz11}. In the case of \citet{wz11}, the 90\% enclosed luminosity radii are adopted to compare with other studies. These histograms are normalized by peak values. The numbers of star clusters found in both studies are presented in each panel. The solid and dashed lines represent zero points and peak positions of differences, respectively. \label{fig:comp.coord.rad}}
	\end{figure}

In addition to \citet{bic08}, \citet{wz11} also presented a catalog of star clusters they found in the MCPS images. They investigated stellar overdensities in LMC fields using stars brighter than 20.5 mag in $V$ band. By both King and Elson-Fall-Freeman model fitting of the surface brightness profiles of each cluster, they determined center coordinates, central surface brightness, tidal radii, and 90\% enclosed luminosity radii of 1,066 LMC star clusters. We compare the center coordinates and radii determined in this study with those given by \citet{bic08} and \citet{wz11} in Figure \ref{fig:comp.coord.rad}.

Figure \ref{fig:comp.coord.rad}(a) -- (f) show the differences of center coordinates of star clusters among three studies. The numbers of star clusters common in both this study and \citet{wz11}, both this study and \citet{bic08}, and both \citet{bic08} and \citet{wz11} are 626, 1,645, and 677, respectively. The center coordinates of star clusters from all three different studies are mostly consistent within 10$\arcsec$. Figure \ref{fig:comp.coord.rad}(g) -- (i) show the differences of radii of star clusters. The numbers of star clusters common in both this study and \citet{wz11}, both this study and \citet{bic08}, and both \citet{bic08} and \citet{wz11} are 615, 1,708, and 681, respectively. \citet{wz11} presented four kinds of radii for each cluster, core radii and 90\% enclosed luminosity radii obtained by fitting using two different models, respectively, and we adopted the 90\% enclosed luminosity radius from King model fitting results as a radius to compare with the results from other studies. The values for the cluster radii derived in this study are in better agreement with \citet{bic08} than with \citet{wz11}.

\subsection{Cluster Sample Selection and Age Estimation}

For this study, we selected 102 star clusters that have red and bright stars in near-infrared $(K_{s} - (J-K_{s}))$ CMDs based on 2MASS catalog \citep{skr06}. These stars are considered as AGB star candidates. The AGB selection method is described in \S4.1. For 96 and 83 of these star clusters, we estimated foreground reddenings and ages, respectively, with reasonable optical photometry as follows.

\begin{figure}[hbt]
\epsscale{1}
\plotone{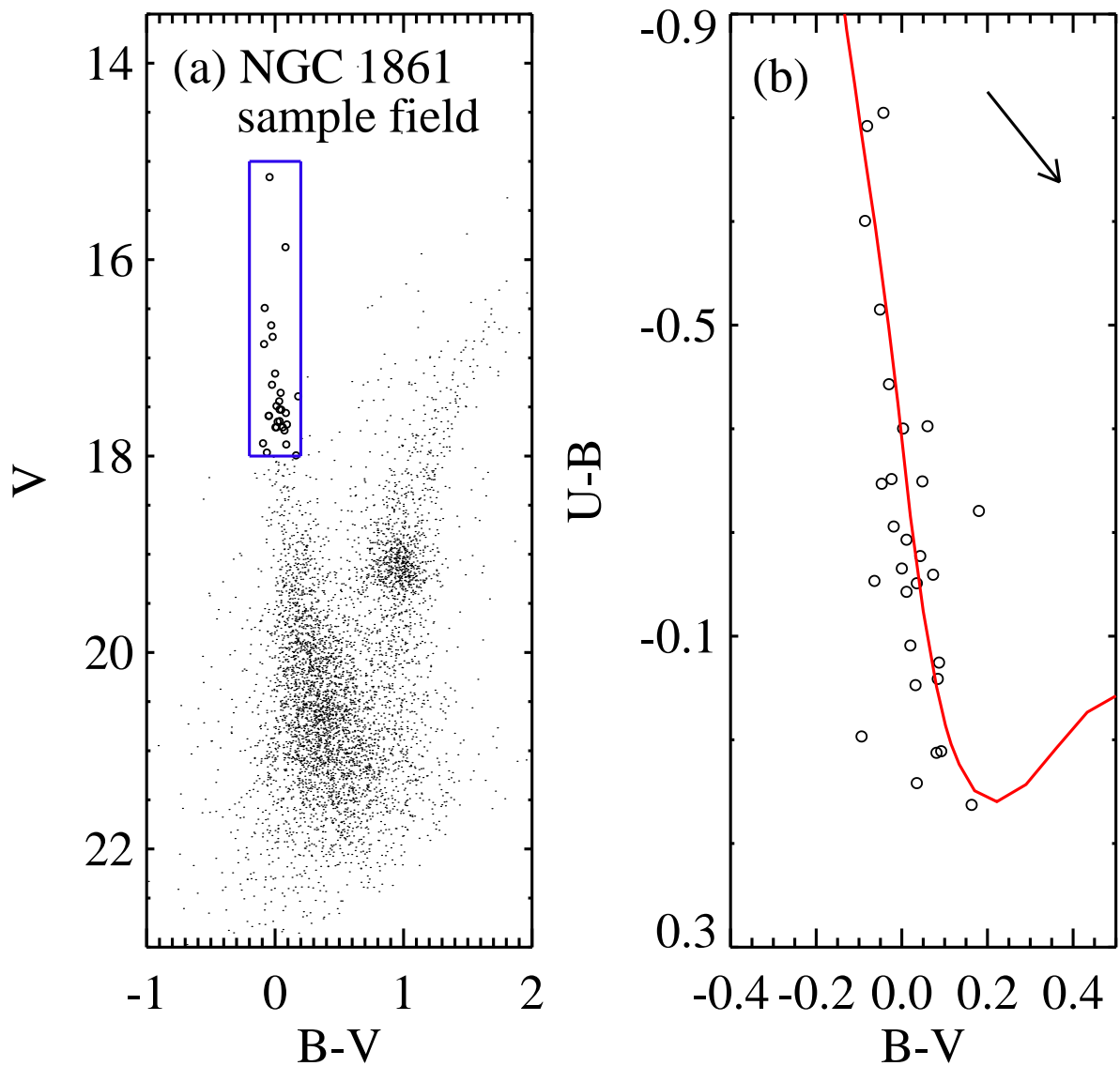}
\caption{(a) $(V-(B-V))$ CMD of stars at 44$\arcsec$ -- 200$\arcsec$ from the center of NGC 1861. The box represents a region where most of stars are field main-sequence stars (open circles). 
(b) $((U-B)-(B-V))$ CCD of field main-sequence stars selected in (a). The solid line is a ZAMS \citep{mar08} shifted according to $E(B-V) = 0.10$ mag. The arrow indicates a direction of reddening vector. \label{fig:red}}
	\end{figure}

We chose member stars of each star cluster using the center coordinates and the radii estimated before (see \S3.1). The value of foreground reddening, $E(B-V)$, was estimated by shifting the zero age main-sequence (ZAMS) in the $((U-B)-(B-V))$ color-color diagram (CCD). 
We used the bright main-sequence stars ($V \lesssim 18 $) located in the outer region up to 200 -- 300$\arcsec$ of each star cluster to obtain the CCDs, and compared the sequence of stars in the CCDs with the ZAMS in the Padova models \citep{mar08} as in Figure \ref{fig:red}(b). The errors of $E(B-V)$ values are about 0.02 mag typically.

\begin{figure}[hbt]
\epsscale{1}
\plotone{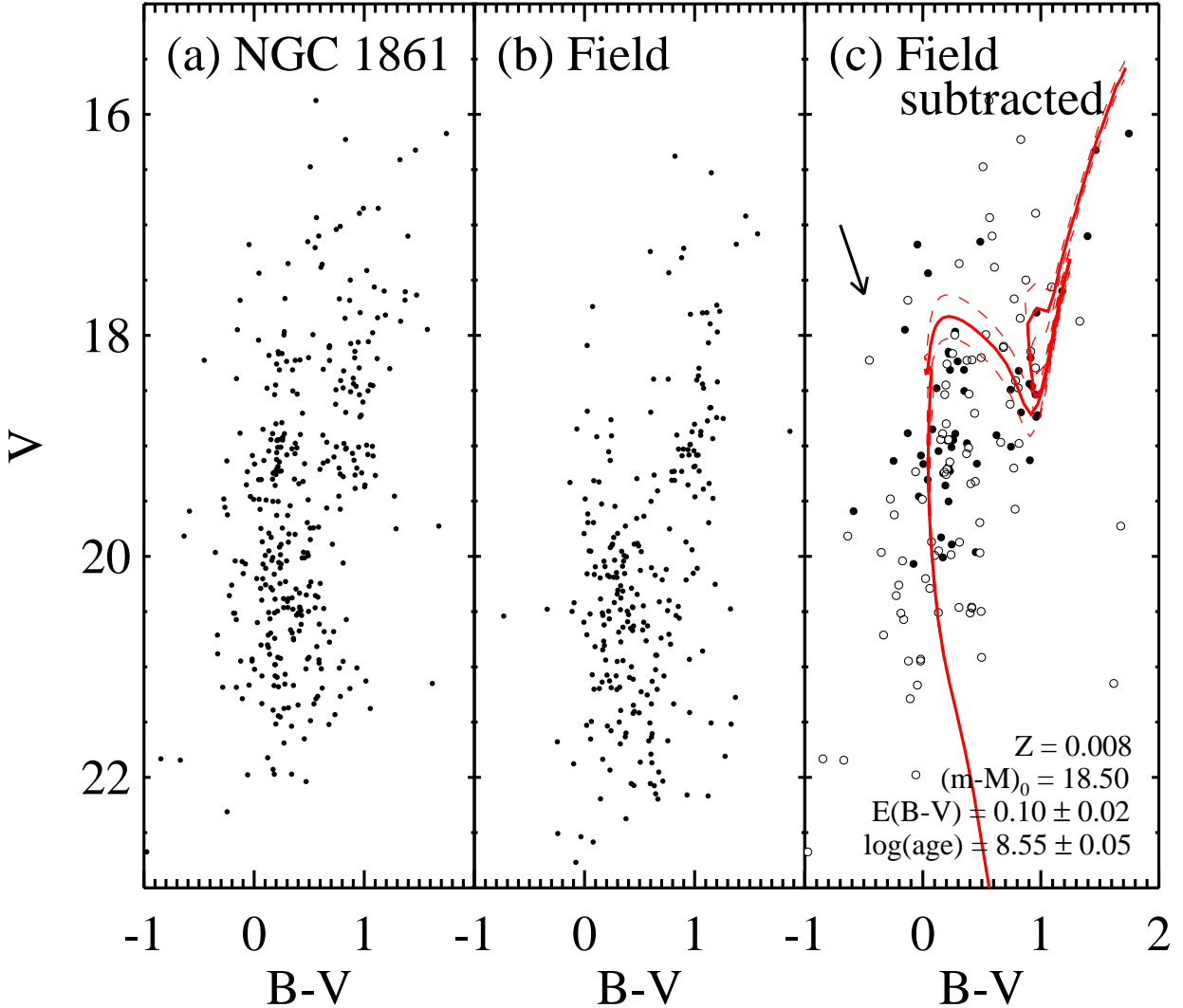}
\caption{(a) $(V-(B-V))$ CMD of stars in NGC 1861 region. (b) $(V-(B-V))$ CMD of stars in the field region around NGC 1861 with the same area as the cluster region. 
(c) $(V-(B-V))$ CMD of stars in NGC 1861 after statistical subtraction of field stars. The filled and open circles represent the stars with the photometric errors of $err(B-V) \leq$ 0.1 mag and $>$ 0.1 mag, respectively. The solid line is a Padova isochrone for log(age [yr]) = 8.55 and $Z = 0.008$ \citep{mar08}. It is shifted according to $E(B-V) = 0.10$ mag and $(m-M)_0 = 18.50$ mag, and the two dashed lines represent isochrones for log(age[yr]) = 8.5 and 8.6, which tell errors of ages. The arrow indicates a direction of reddening vector. The values of physical parameters of NGC 1861 (metallicity, distance modulus, reddening, and age) are shown in the right lower corner. \label{fig:stat.cmd}}
	\end{figure}

Figure \ref{fig:stat.cmd}(a) and (b) show $(V-(B-V))$ CMDs of both a cluster and a field region for NGC 1861, as an example. The cluster CMD is expected to contain field stars as well as cluster stars. In order to minimize the field contamination, we performed statistical subtraction of field CMDs for cluster CMDs. We counted the number of stars in cluster CMDs with that of stars in field CMDs for the same area as the cluster area for each color and magnitude bin ($\Delta(B-V) \sim 0.5$ and $\Delta V \sim$ 1), and subtracted statistically field stars from cluster stars for each bin. 
Figure \ref{fig:stat.cmd}(c) displays a field-subtracted CMD of NGC 1861. It shows a clear stellar sequence with a smaller number of stars than the original CMD. 
We determined ages of 83 star clusters by isochrone fitting in the field-subtracted $(V-(B-V))$ CMDs, assuming the distance modulus $\mathrm{(m-M)_0}$ = 18.50 mag and $Z = 0.008.$ Cluster ages were derived using isochrones of \citet{mar08}.
We used only the stars with small errors of colors with $err(B-V) < 0.1$ mag for isochrone fitting.
The ages of the other 19 star clusters could not be derived, because they were not covered by the MCPS observation fields or are older than 1 Gyr.
In this case, we adopted the ages of these clusters from the literature (\citealt{ef85}, \citealt{gir95}, \citealt{pu00}, \citealt{gou11}, \citealt{pop12}).
Table \ref{tab:sc.prop} lists physical parameters of 102 star clusters that will be used for AGB star selection.

\begin{figure}[hbt]
\epsscale{1}
\plotone{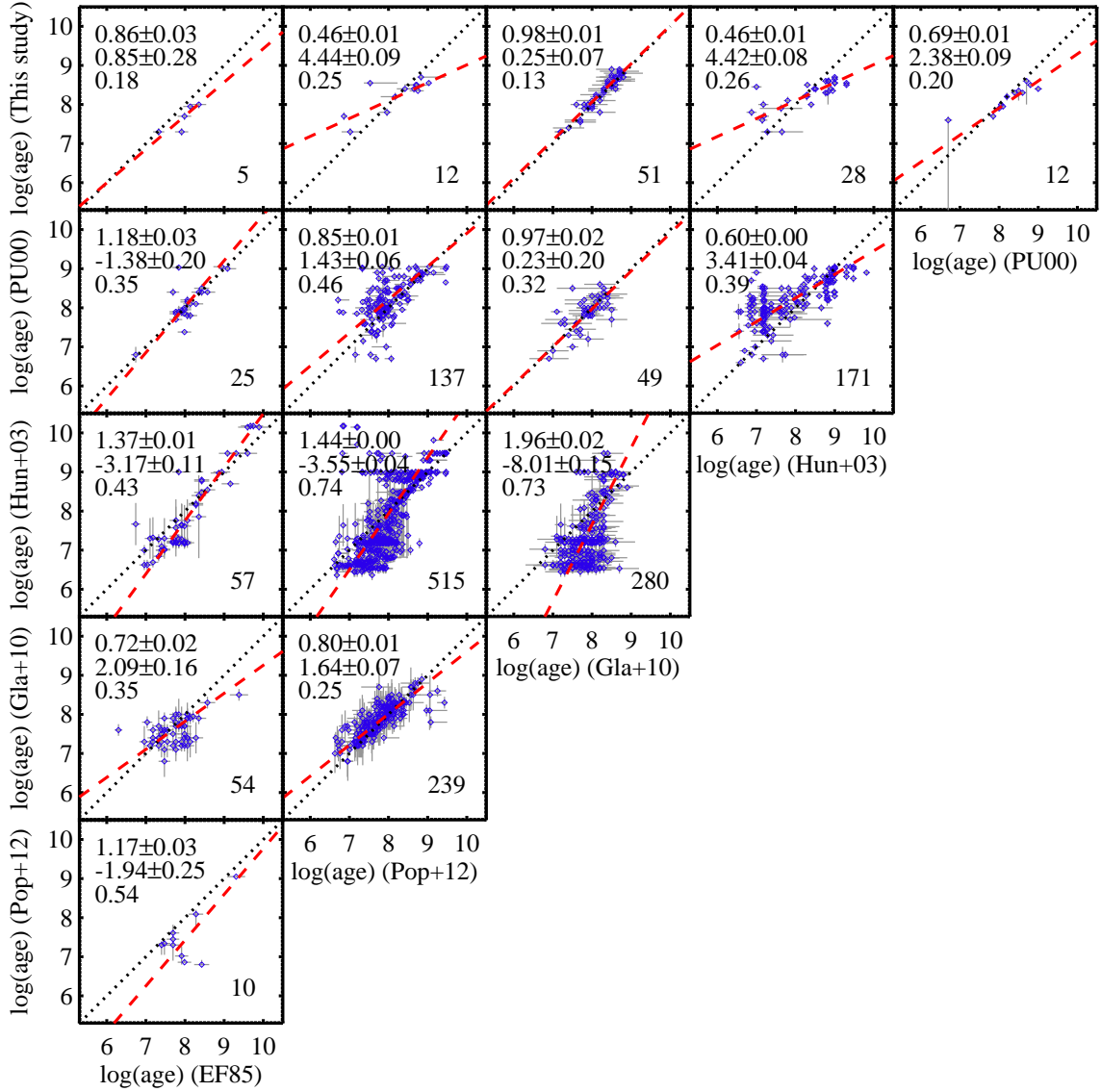}
\caption{Comparison of the ages determined by various references (\citealt{ef85}; EF85, \citealt{pu00}; PU00, \citealt{hun03}; Hun+03, \citealt{gla10}; Gla+10, \citealt{pop12}; Pop+12). The dotted lines are one-to-one relations. The dashed lines represent linear fitting results, and their slopes, zero points, and the root mean square errors are shown in the left upper corner of each panel. 
\label{fig:comp.age}}
	\end{figure}

We compared the ages derived in this study with those given in other references, \citet{ef85}, \citet{pu00}, \citet{hun03}, \citet{gla10}, and \citet{pop12}, as shown in Figure \ref{fig:comp.age}.
The isochrone fitting method used for the resolved stars in star clusters is considered to be more reliable than other age-dating methods based on the integrated color or spectra of the star clusters. 
\citet{pu00}, \citet{gla10}, and this study used the isochrone fitting method,  
while \citet{ef85}, \citet{hun03}, and \citet{pop12} analyzed the integrated color of the star clusters. \citet{ef85} and \citet{hun03} analyzed $(U-B)$ and $(B-V)$ colors of the star clusters, and \citet{pop12} performed a Monte Carlo simulation with their own star cluster simulation software (MASSive CLuster Evolution and ANalysis, MASSCLEAN; \citealt{ph10}) to reproduce the cluster colors, $(U-B)_0$ and $(B-V)_0$. 
The ages based on the isochrone fitting method show good agreement with each other. They also show correlations with the ages based on integrated colors, but with significant scatters and non-unity slopes. It is noted that the ages derived from integrated colors by \citet{pop12} show a better correlation with the isochrone-fitting ages, compared with other results in \citet{ef85} and \citet{hun03}. 

\begin{figure}[hbt]
\epsscale{1}
\plotone{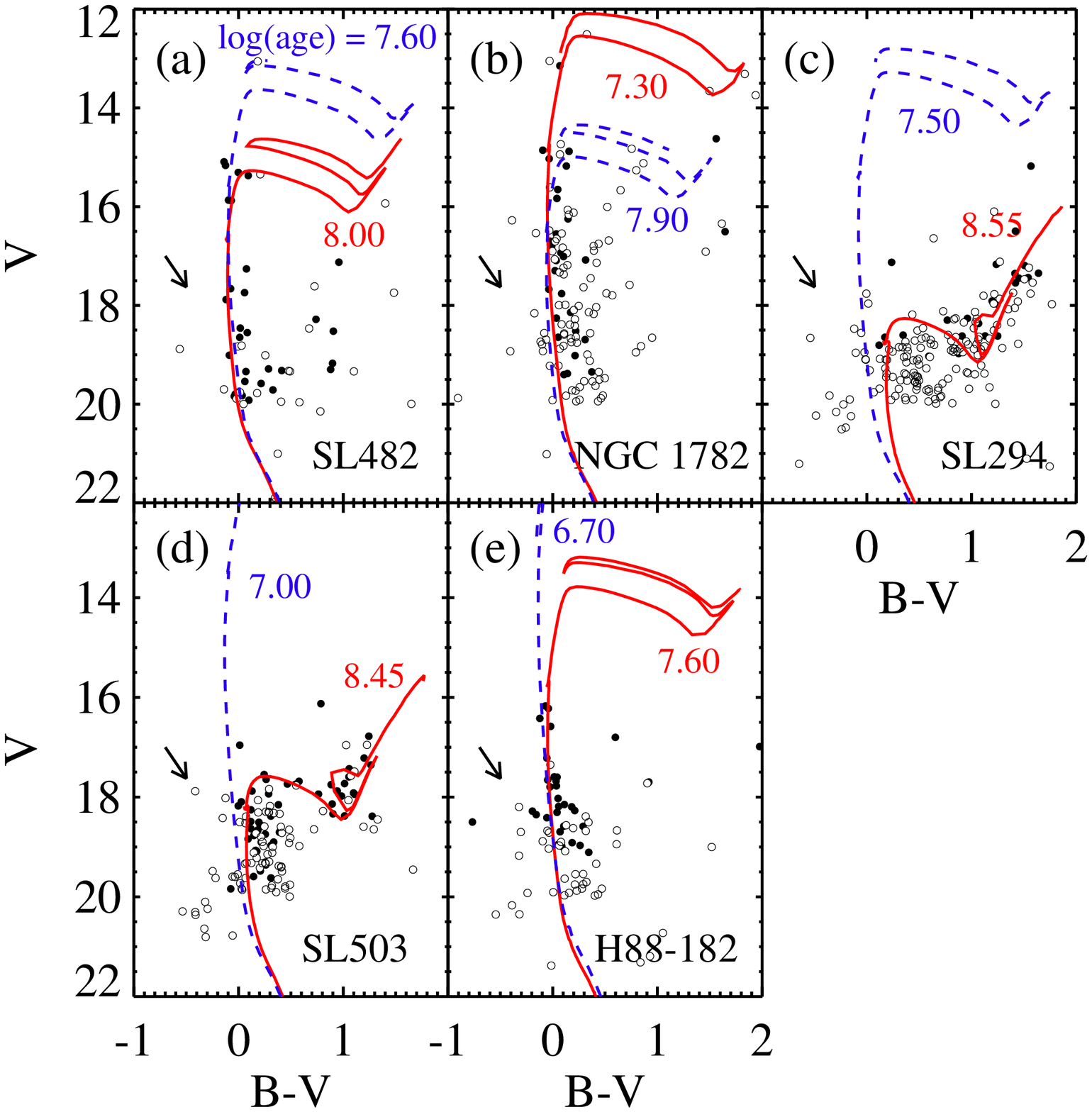}
\caption{$(V-(B-V))$ CMDs and isochrone fitting results of the star clusters that show the largest difference of ages from other references: (a) SL482 \citep{gla10}, (b) NGC 1782 \citep{ef85}, (c) SL294 \citep{pop12}, (d) SL503 \citep{hun03}, and (e) H88-182 \citep{pu00}. 
The filled and open circles represent the same as in Figure \ref{fig:stat.cmd}(c).
The solid lines and dashed lines represent Padova isochrones ($Z = 0.008$) for the ages  determined in this study and other studies, respectively. These are shifted according to the determined foreground reddening and $(m-M)_0 = 18.50$ mag.  The arrow indicates a direction of reddening vector. \label{fig:comp.cmd}}
	\end{figure}

Figure \ref{fig:comp.cmd} shows the CMDs of five star clusters that have the largest difference of ages between this study and others: 
SL482 \citep{gla10}, NGC 1782 \citep{ef85}, SL294 \citep{pop12}, SL503 \citep{hun03}, 
and H88-182 \citep{pu00}. We plotted the Padova isochrones for $Z=0.008$ and the ages
corresponding to the values derived in this study and other references.
In the case of SL482, NGC 1782, and H88-182, it is difficult to determine their ages reliably because the number of the brightest stars around the main-sequence turnoff is small. 
However, in the case of other two star clusters, SL294 and SL503, their CMDs are not matched by the isochrones for the ages derived by \citet{pop12} and \citet{hun03} respectively, while they are by the isochrones for the ages derived in this study.

\section{Results}

	\subsection{Selection of AGB Stars}

While it is difficult to distinguish AGB stars from RGB stars in optical CMDs, it is relatively easier in near-infrared CMDs. 
Therefore, we plotted the near-infrared $(K_{s,0}-(J-K_{s})_0 )$ CMDs of resolved stars in 102 star clusters using the 2MASS point source catalog \citep{skr06}, and selected AGB stars using two criteria: (1) stars brighter than the tip of the RGB (TRGB); $K_{s,0} < 12.3$, and (2) stars redder than the RGB; $(J-K_{s})_{0} > -0.075 \times K_{s,0} + 1.85$. We adopted the TRGB magnitude, $K_{s} =$ 12.3 $\pm$ 0.1 estimated from 2MASS point sources in the LMC region \citep{nw00}. The second criterion was suggested from the analysis of the 2MASS photometry for LMC stars by \citet{cio06}. There is no constraint for the brightest and reddest stars.
We use the extinction law in \citet{car89} to derive extinction values for each band, 
adopting $R_V = 3.1$. 

Figure \ref{fig:cum.nir}(b) and (d) show the combined $(K_{s,0}-{(J-K_{s})_0})$ CMDs of all stars in 41 young star clusters with log(age) $< 8.5$ and 61 old star clusters with log(age) $\geq 8.5$, respectively.
We selected the AGB stars brighter and redder than AGB boundary lines as described above (dashed lines).
We also plotted other AGB boundaries used in the references in Figure \ref{fig:cum.nir}(b) and (d): dot-dashed lines used by \citet{muc06} and gray shaded regions used by \citet{nw00}. 
The AGB boundary used in this study is working in the similar way to that in \citet{nw00}.
However, the AGB boundary used by \citet{muc06} includes a significant number of bluer stars, compared with that in this study. This difference will be discussed in \S5.1.

	\subsection{Measurement of $K_s$ Band Luminosity Fraction of the AGB Population in Star Clusters}

We estimated the $K_s$ band luminosity of star clusters by aperture photometry using 2MASS $K_s$ band images. Photometry was performed with IRAF\footnote{IRAF is distributed by the National Optical Astronomy Observatories, which are operated by the Association of Universities for Research in Astronomy, Inc., under cooperative agreement with the National Science Foundation.} APPHOT package. 
For background estimation we used annular apertures with the inner radius to be the sum of the star cluster radius and its error and  the aperture width of 10$\arcsec$. The background level was estimated using the mode value of the annulus region. For cluster photometry we used circular apertures with radius that is the same as the cluster radius. 
Figure \ref{fig:aperture} displays a $K_s$ band image of an example star cluster, NGC 1861, showing the position of apertures for the cluster and background area. 
We calculated the $K_s$ band luminosity of AGB stars in each cluster by summing their luminosity converted from 2MASS magnitudes, and obtained the $K_s$ band luminosity fraction of AGB stars in the star clusters.

Table \ref{tab:agb.lf} summarizes the properties of AGB populations as well as other basic parameters of 102 LMC star clusters.
It lists the number of AGB stars, and the $K_s$ band luminosity and luminosity fraction of AGB stars in each cluster, as well as $K_s$ band integrated magnitude and luminosity of the star clusters.  The magnitudes are based on 2MASS magnitude system. 
We adopted the $K_s$ band absolute magnitude $\mathrm{M_{\odot}^{K_s}} = 3.27$ mag for the Sun, using $\mathrm{M_{\odot}^{V}} = 3.83$ mag \citep{all76} and $(V-K_{s}) = 1.56$ mag \citep{cas12}, to convert the magnitude into the luminosity.

The error of the AGB luminosity fraction is contributed by various errors in addition to the photometric ones estimated for the integrated magnitude of star clusters and presented for the 2MASS magnitude of AGB stars. 
First, there exist the errors of center coordinates and radii of star clusters, influencing the estimation of their integrated luminosity. We performed aperture photometry for all sample clusters, setting different center positions and radii according to their errors.
We selected the upper and lower errors of the integrated luminosity of star clusters, also considering photometric errors. 
Secondly, there are the errors for the luminosity of AGB stars, which consist of (a) 2MASS photometric error for individual AGB stars and (b) the uncertainty from the Poisson error for the number of AGB stars in each star cluster associated with AGB star counts. 
Of these two, the latter is much larger than the former. 
The AGB luminosity error from Poisson error is calculated by multiplying the square root of the number of AGB stars by the mean luminosity for an AGB star in each cluster. We tabulated upper and lower errors of the $K_s$ band luminosity fraction obtained by error propagation, considering overall error budget mentioned above (see Table \ref{tab:agb.lf}). Note that the smaller the numbers of AGB stars in star clusters are, the larger the errors of the luminosity fraction are.

Additionally, we checked the field contamination in AGB star count. We investigated the outer region of each star cluster with the radius five times larger than the cluster radius. The number of the AGB stars in this field region is normalized to the cluster area.
We calculated the amount of field contamination by multiplying the mean luminosity of AGB stars in star clusters by the number of the normalized field AGB stars, and subtracted it from the luminosity of AGB stars in star clusters.
However, the effect of the field contamination to estimating the luminosity fraction of AGB stars is much smaller than that of the Poisson error of the AGB star count. These field-subtracted values are also listed in Table \ref{tab:agb.lf}.

Figure \ref{fig:main1}(a) shows the $K_s$ band luminosity fraction of AGB stars in our sample clusters as a function of ages. 
Figure \ref{fig:main1}(b) displays the mean values of the $K_s$ band luminosity fraction of AGB stars in the star clusters. 
These values are also listed in Table \ref{tab:comp.lf} including the field-subtracted values. 
The mean values represent the ratios of the total $K_s$ band luminosity of AGB stars in the star clusters to the total $K_s$ band luminosity of the star clusters for each age bin. The error of mean values is calculated from the individual errors of the star clusters by error propagation. 
We also derived a sequence to represent approximately the sequence of bright clusters, as plotted in the figure. The bright clusters have the luminosity of L$_{K_s} > 8.5 \times 10^4$ L$_{\odot}.$
Following features are noted in Figure \ref{fig:main1}.
First, the mean values and the bright cluster sequence increase, as log(age) increases from 8.0 to 8.5, reaching 0.6 and 0.8 at log(age) $\sim 8.5$, respectively. 
They decrease thereafter.
Second, there is a large scatter in the mean values for given ages. 
We discuss these features in \S5.2.2 in detail.

\section{Discussion}

\subsection{Comparison with Previous Studies}

In Figure \ref{fig:main1}(b), we also plotted the results for 19 LMC star clusters given by \citet{muc06} for comparison. 
There are 12 star clusters common between this study and \citet{muc06}.
We determined foreground reddenings of six star clusters (NGC 1806, NGC 1866, NGC 1987, NGC 2108, NGC 2134, and NGC 2136) by using the method mentioned above (see \S3.2). 
For the other three star clusters(NGC 1831, NGC 2173, and NGC 2249), the field stars were not covered by the MCPS. In this case, we adopted foreground reddening values from the optical reddening map of the MCs given by \citet{has11}. \citet{has11} presented the optical reddening map of the MCs obtained by comparing the theoretical color of the red clump with its observed one based on the OGLE III data. The reddening values of the others (NGC 2162, NGC 2190, and NGC 2231) could not be determined neither in this study nor in \citet{has11}. We assumed the typical extinction value, $E(B-V) = 0.1$, for these three clusters.
For ages, we determined ages of three of these clusters (NGC 1866, NGC 2134, and NGC 2136). The ages of the others are adopted from \citet{ef85} and \citet{gir95}, because we could not use the MCPS catalog for these clusters.

\begin{figure}[hbt]
\epsscale{1}
\plotone{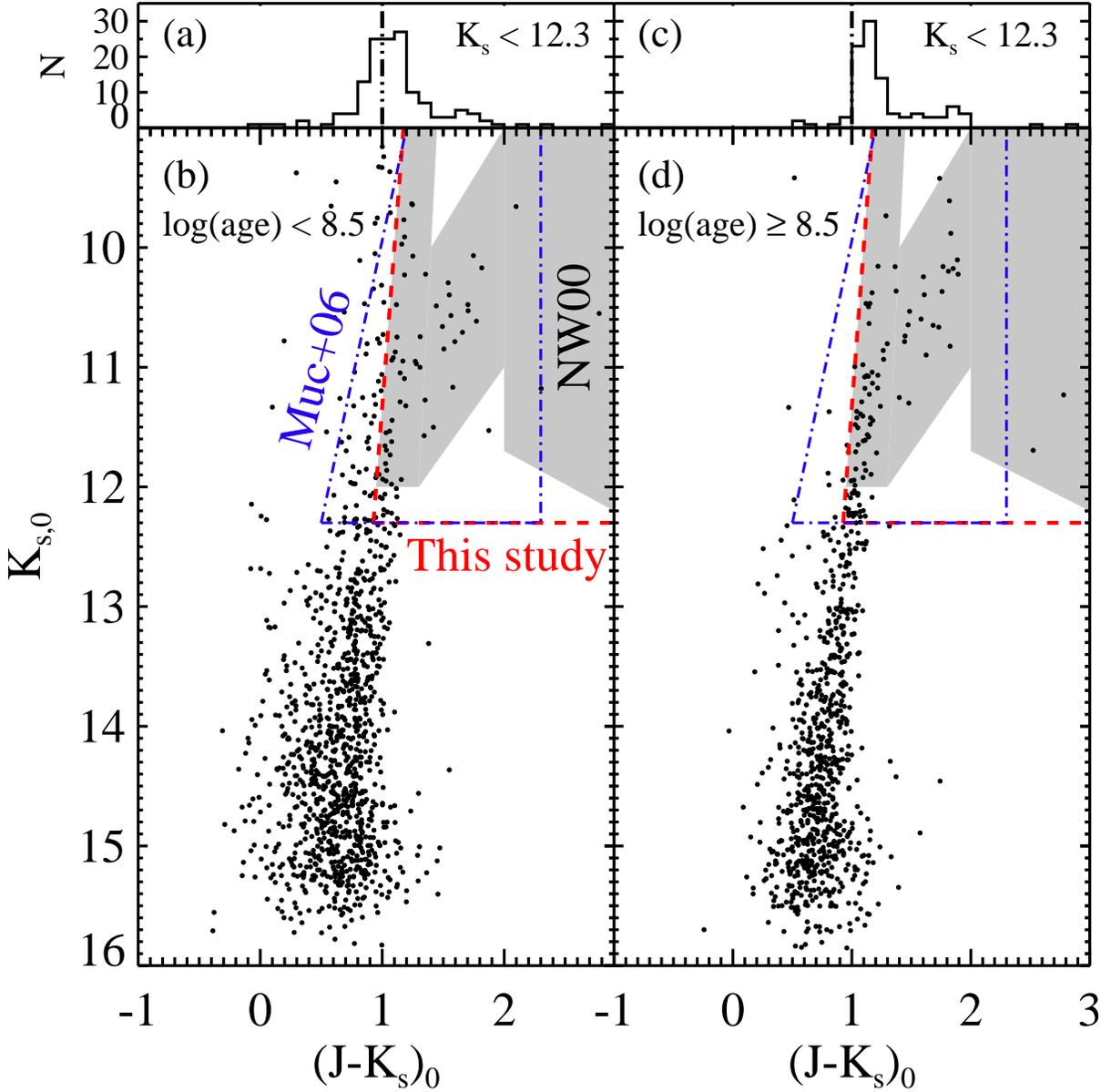}
\caption{AGB selection criteria. (a) The color distribution of the bright stars brighter with $K_s < 12.3$ in 41 young star clusters with log(age) $<$ 8.5. (b) A combined and dereddened $(K_{s,0}-(J-K_{s})_0)$ CMD for star clusters with log(age) $<$ 8.5. The dashed lines and the dot-dashed lines indicate criteria to distinguish AGB stars from other populations in this study and \citet{muc06}, respectively. The shaded area represents the area where \citet{nw00} suggested that AGB stars are concentrated. (c) and (d) the same as (a) and (b) for 61 star clusters with log(age) $\geq$ 8.5. \label{fig:cum.nir}}
	\end{figure}

\begin{figure}[hbt]
\epsscale{1}
\plotone{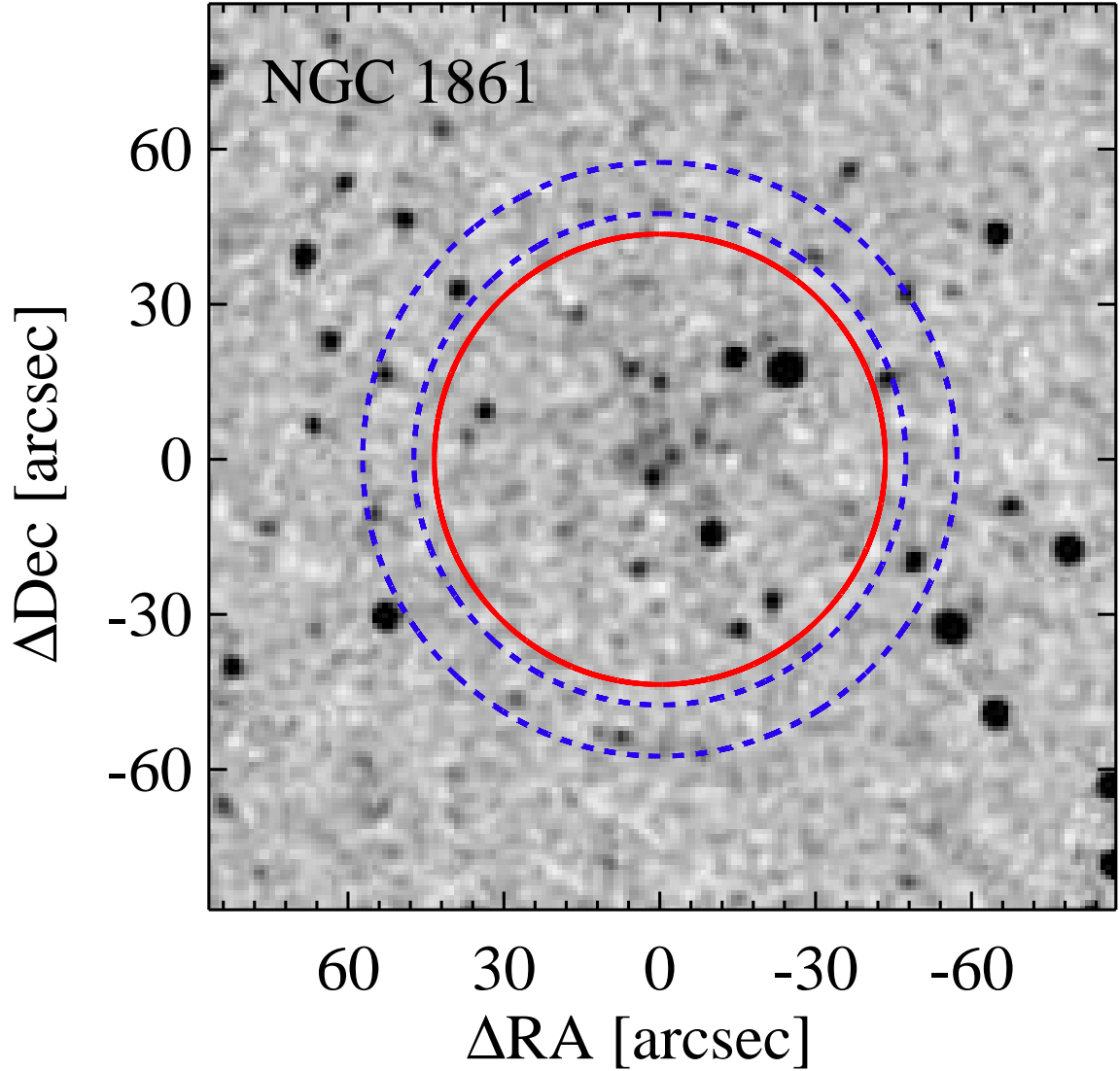}
\caption{A gray scale map of a 2MASS $K_s$ band image of NGC 1861. The solid line circle indicates an aperture with radius 44$\arcsec$ for cluster photometry. The two dashed line circles represent the annuli with radii 48$\arcsec$ and 58$\arcsec$, used for background estimation. \label{fig:aperture}}
	\end{figure}

\begin{figure}[hbt]
\epsscale{0.9}
\plotone{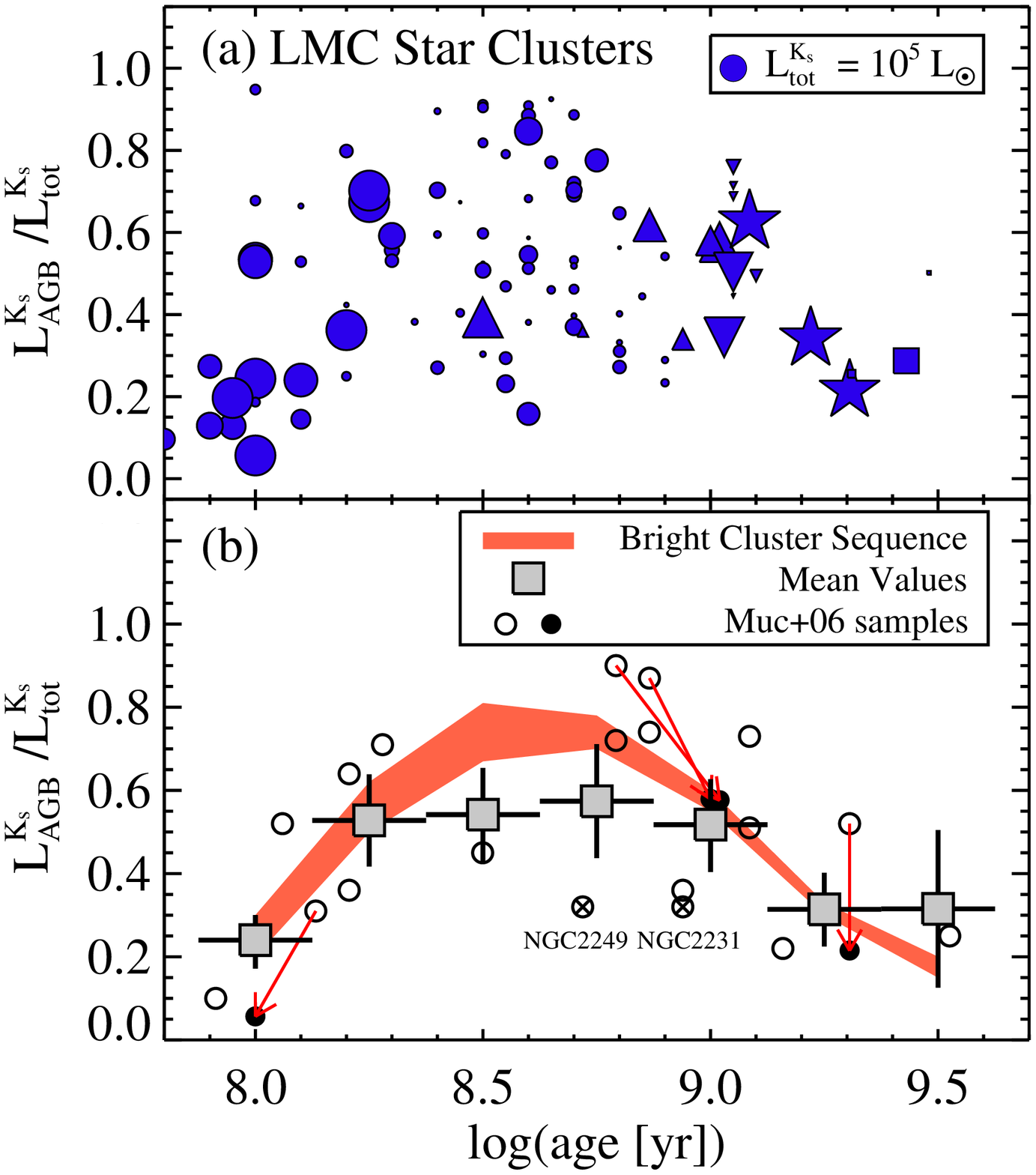}
\caption{(a) The $K_s$ band luminosity fraction of AGB stars in 102 LMC star clusters as a function of ages. The circles and other symbols represent the star clusters of which ages were determined in this study and other studies, respectively (triangles: \citealt{ef85} and \citealt{gir95}, stars: \citealt{gou11}, upside-down triangles: \citealt{pu00}, and squares: \citealt{pop12}). The symbol size reflects the integrated luminosity of each cluster.
(b) Mean values in each age bin (squares) and bright cluster sequence (shaded region) of the $K_s$ band luminosity fraction of AGB stars in star clusters. The lower boundary of the shaded region show the result after field-subtraction, and for mean values, the field-subtracted results are within error bars.
The error bars for mean values indicate the errors estimated by error propagation. The open circles represent the result from \citet{muc06} including NGC 2249 and NGC2231 with L$_{K_s} < 10^5$ L$_\odot$ (cross marks). The filled circles indicate the result derived in this study for the star clusters that show a large discrepancy in the AGB luminosity contribution. \label{fig:main1}}
	\end{figure}

\begin{figure}[hbt]
\epsscale{1}
\plotone{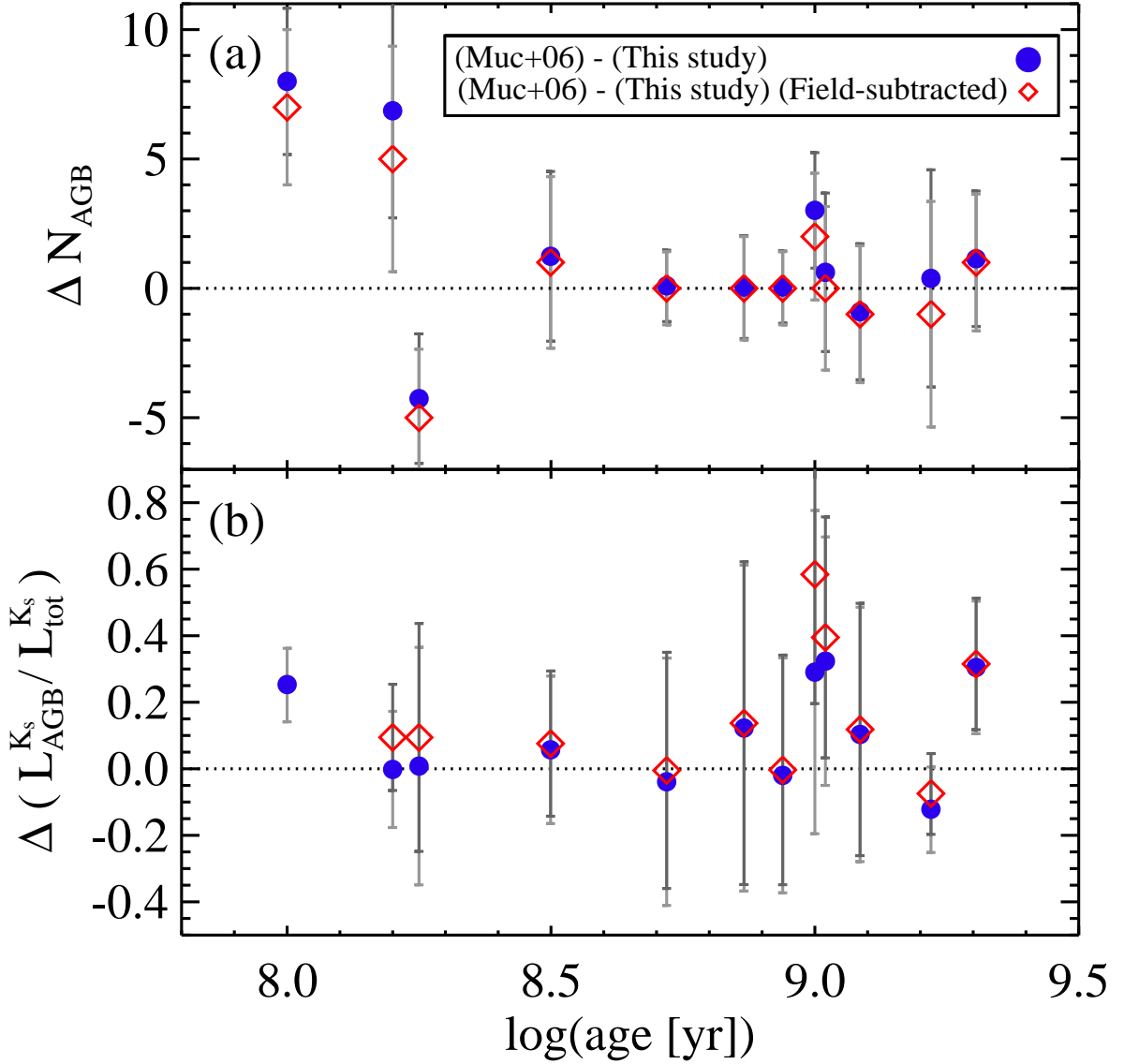}
\caption{(a) The difference of the number of AGB stars in 12 star clusters between \citet{muc06} and this study (filled circles). The open diamonds represent the field-subtracted results. The error bars indicate Poisson errors for the number of AGB stars. (b) Same as (a) but for the $K_s$ band luminosity fraction of AGB stars in star clusters. The error bars indicate the errors estimated by error propagation. \label{fig:comp.muc06}}
	\end{figure}

We also noticed that the AGB selection criteria of \citet{muc06} are different from those of this study (see Figure \ref{fig:cum.nir}). \citet{muc06} included as AGB candidates the blue stars that are excluded in this study, but the color and magnitude of these stars indicate that they are  bright RGB and supergiant populations according to the analysis of \citet{nw00}. 
The color distribution of the stars brighter than $K_s$ = 12.3 for young star clusters with log(age) $<$ 8.5 shows a blue excess at $(J-K_s )_0 < 1.0$ (see Figure \ref{fig:cum.nir}(a)), while little blue excess is seen for the case of old star clusters with log(age) $>$ 8.5 (see Figure \ref{fig:cum.nir}(c)). However, those blue stars used in the analysis of \citet{muc06} do not significantly influence the luminosity fraction of AGB stars in the star clusters, because they are somewhat faint (see Figure \ref{fig:comp.muc06}).
In addition, there are three red stars not included in the AGB boundary of \citet{muc06} but contained in that of this study, which can be dusty AGB stars. We found that these stars affect little our results.

The number of AGB stars shows a large discrepancy between our results and \citet{muc06} for the young star clusters as shown in Figure \ref{fig:comp.muc06}(a), but the difference in the luminosity fraction of AGB stars in the star clusters is much smaller as shown in Figure \ref{fig:comp.muc06}(b). 

In Figure \ref{fig:main1}(b), the results for 19 star clusters from \citet{muc06} are included, and those derived in this study are also plotted for four star clusters that show a large discrepancy ($>$ 0.2) in the AGB luminosity fraction (see Figure \ref{fig:comp.muc06}(b)). Other eight clusters show the consistent results with \citet{muc06}.
Most of these clusters studied by \citet{muc06} are brighter than L$_{K_s} = 10^5$ L$_\odot$, except for two star clusters, NGC 2249 and NGC 2231. 
Indeed they are following the bright cluster sequence derived in this study, with some scatter.


	\subsection{Comparison with SSP Models}
	
		\subsubsection{Mock Cluster Experiments with SSP models}
	
We compare our results with the expectation of five theoretical models:
(1) an EPS model of \citet{mar05} (called Maraston model), 
(2) a model based on the isochrone set of \citet{gir02} (called Padova02 model), 
(3) a model based on the improved isochrone set of \citet{mar08} corrected by \citet{gir10} (called Padova10 model),
(4) a model based on the isochrone set of Pietrinferni et al. (2004, 2006) assuming a given overshooting efficiency (called BaSTI\_os model), 
and (5) the same as (4), but for null overshooting efficiency (called BaSTI\_std model).
The Maraston model calculates the luminosity contribution of AGB stars directly using the fuel consumption approach. 
\citet{gir02} and \citet{mar08} corrected by \citet{gir10} presented theoretical stellar evolutionary tracks and isochrones used in various EPS models.
The detailed information of these models is described as follows.

The Maraston model calculated the amount of the fuel consumption of each population in the post-main-sequence phases \citep{rb86} directly, and converted it to observables assuming the effective temperature and surface gravity for evolutionary stages. We obtained the Maraston model prediction for the luminosity contribution of AGB stars in SSPs as a function of ages \citep{muc06}. 

The theoretical isochrones of \citet{gir02} are based on isochrones of \citet{gir00} for low- and intermediate-mass stars (M $\leq 7$ M$_\odot$) and \citet{ber94} for high-mass stars (M $> 7$ M$_\odot$). Note that stellar evolutionary tracks of \citet{gir00} are adopted in the BC03 model. In this model, they included the simplified TP-AGB phase and no circumstellar dust that form in AGB stars.

\citet{mar08} presented optical to far-infrared isochrones with improved TP-AGB models. \citet{gir10} searched for the TP-AGB population of 12 galaxies in the ACS Nearby Galaxy Survey Treasury, and found that the model of \citet{mar08} creates more TP-AGB populations than observed ones. They corrected the mass-loss rate of TP-AGB stars in this model by reducing their lifetime (\citealt{bw91}, \citealt{wil00}). This model reproduces the number of TP-AGB stars well, but still has uncertainties in their 1.6$\mu$m band flux contribution \citep{mel12}.
Additionally, we adopted the isochrone set that includes the circumstellar dust from AGB stars 
\citep{gro06} and assumed that the dust composition is 100\% silicate and 100\% amorphous carbonate dust for O-rich and C-rich AGB stars, respectively. 

These two kinds of Padova isochrones consider the core convective overshooting. In order to investigate this effect, we used the BaSTI isochrone dataset (Pietrinferni et al. 2004, 2006) without overshooting efficiency. We selected a scaled solar isochrone set that contains the extended AGB population. 
We calculated the predicted $K_s$ band luminosity fraction of AGB stars in SSPs by analyzing the mock star clusters produced from the isochrones of \citet{gir02}, \citet{mar08}, and \citet{pie04}. We used the isochrone sets with metallicity  Z = 0.008, corresponding to [Z/H] $\sim$ -0.35 for mock cluster experiments. This value is close to the mean value for the LMC, for our cluster samples.  The method is described as follows.

We created model stars in mock star clusters assuming the Salpeter initial mass function, and assigned $J$ and $K_s$ magnitudes to each star using four different isochrone sets.
In addition, we considered magnitude errors for each star in order to make the model prediction more realistic. In the 2MASS catalog, mean magnitude errors and error variances of stars in each magnitude bin vary as a function of magnitudes. 
We assumed that photometric errors of model stars are distributed normally with the mean value and the variance according to magnitude bins. In $(K_{s,0}-(J-K_{s})_{0})$ CMDs of model stars in mock star clusters with empirical magnitude errors, AGB stars are selected with the same criteria as done for observational data (see \S4.1). The $K_s$ band luminosities of mock star clusters and AGB stars are calculated by summing the $K_s$ band luminosity of member stars and AGB stars, respectively. Note that for the BaSTI models, they do not provide the magnitudes of 2MASS filter system, so that we adopted those of the Bessell filter system.
Finally, we obtained the $K_s$ band luminosity fraction of AGB stars in mock star clusters with three different mass ($10^4$, $10^5$, and $10^6$ $\mathrm{M}_\odot$), assuming 15 different ages (8.0 $\leq$ log(age) $\leq$ 9.5) for each mass scale.

		\subsubsection{Stochastic Effect in Estimating Light from AGB Stars}

Stochastic effects are inevitable in AGB star counts because of the short lifetime of AGB stars. In Figure \ref{fig:main1}(a), there is a large scatter in the $K_s$ band luminosity fraction of AGB stars in the star clusters derived in this study. The AGB stars evolve fast, and become blue and faint when they enter the post-AGB phase. Therefore, we cannot detect all stars that have entered the AGB phase. As a result it makes the luminosity fraction of AGB stars in star clusters lower than expected.
Especially, this effect becomes significant for faint star clusters, because of a smaller number of stars. Bright star clusters ($\mathrm{L_{tot}^{K_s} \gtrsim 8.5 \times 10^4 L_\odot}$) show relatively smaller scatters and seem to be located along a sequence. 

\begin{figure}[hbt]
\epsscale{1}
\plotone{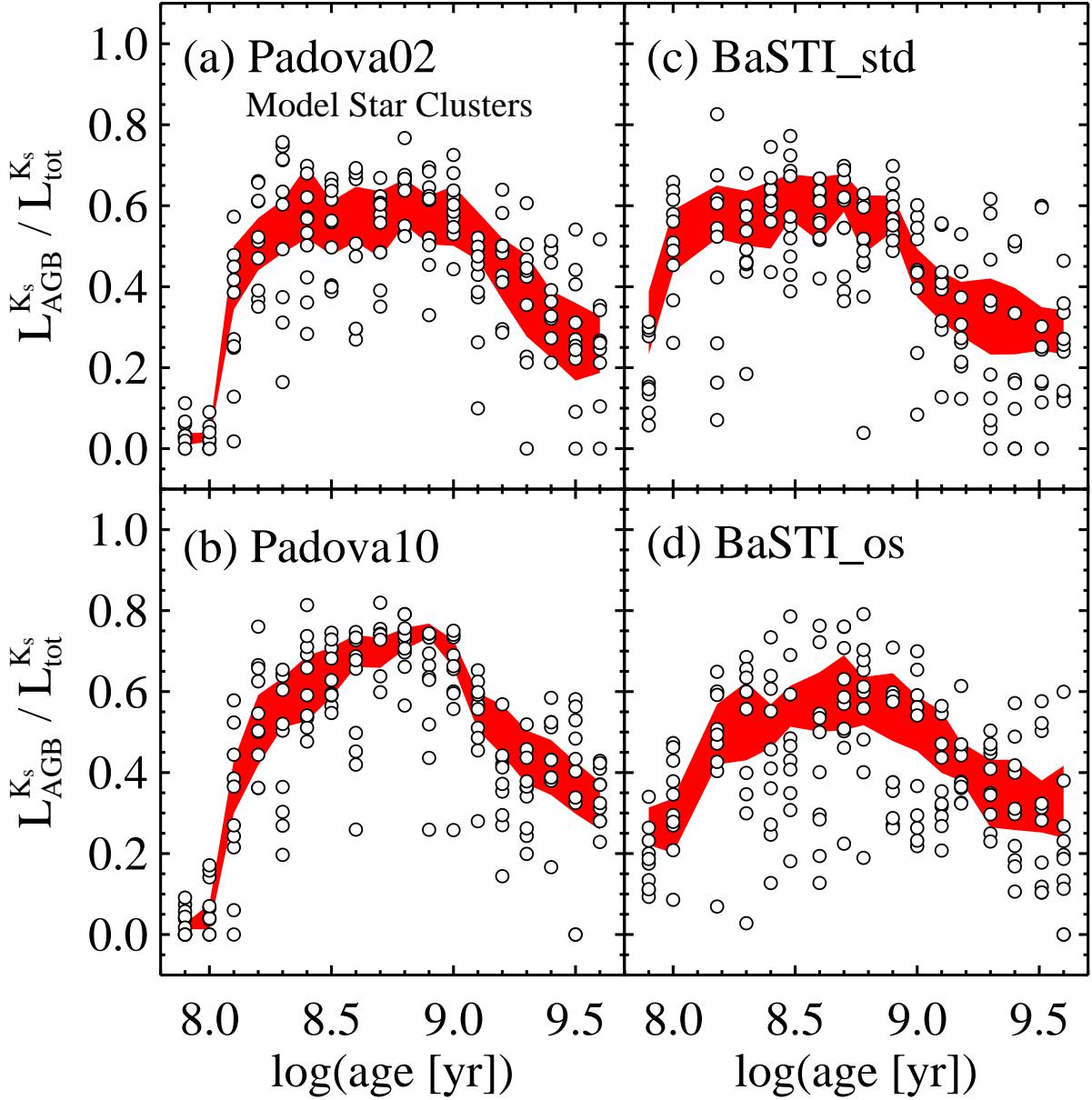}
\caption{The $K_s$ band luminosity fraction of AGB stars in mock star clusters using four different isochrone sets: (a) Padova02, (b) Padova10, (c) BaSTI\_std, and (d) BaSTI\_os models. The shaded region indicates the sequence of the massive mock clusters with M = $10^6$ M$_\odot$. The circles represent the mock star clusters with M = $10^5$ M$_\odot$. Note that the scatter in the AGB luminosity fraction in SSPs is smaller in the Padova10 model than other models. The Padova10 model creates more AGB stars than other models, so that the stochastic fluctuation can be less. \label{fig:stoch}}
	\end{figure}

This trend also appears in the mock star clusters (see Figure \ref{fig:stoch}). We made mock star clusters with 15 different ages and three different masses using four different isochrone sets as above. For each age and mass bin, we made 10 mock star clusters in order to analyze them statistically. Figure \ref{fig:stoch} shows the $K_s$ band luminosity fraction of AGB stars in the mock star clusters as a function of ages. In this figure, the most massive mock star clusters with M = $10^6$ M$_\odot$ show the smallest scatter among mock clusters with other mass scale, and make a well-defined sequence (shaded region). However, mock star clusters with M = $10^5$ M$_\odot$ (circles) are distributed with large scatter from the massive mock cluster sequence.
Therefore we presented two representatives for the observational data: mean values and the bright cluster sequence.

		\subsubsection{Comparison with Model Expectation}
	
\begin{figure}[hbt]
\epsscale{1}
\plotone{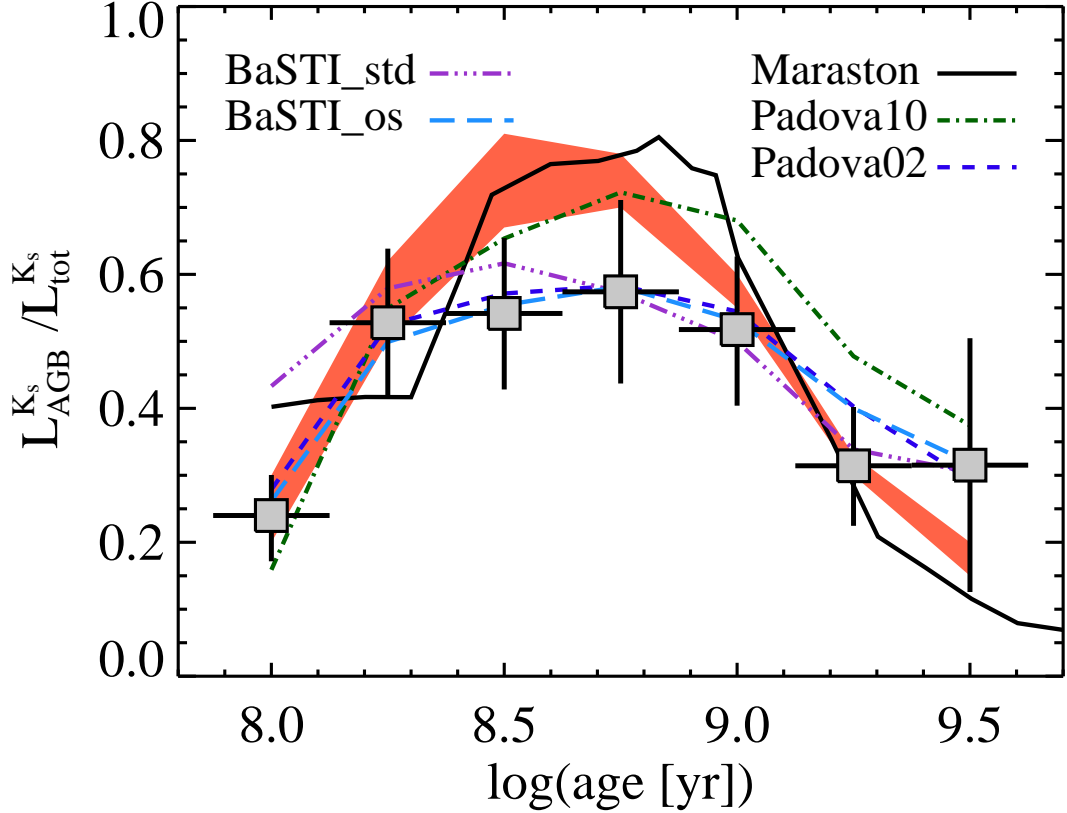}
\caption{Comparison of five SSP model predictions with our results. Squares and shaded region are the same as in Figure \ref{fig:main1}(b). The solid line represents the Maraston model expectation for both the E-AGB and TP-AGB computed at [Z/H] = -- 0.33, and the dashed line and the dot-dashed line indicate respectively Padova02 and Padova10 model expectations from massive mock star clusters with M = $10^6$ M$_\odot$. The BaSTI\_std (triple-dot dashed line) and BaSTI\_os (long dashed line) model predictions for massive clusters are also plotted. \label{fig:main2}}
	\end{figure}

Figure \ref{fig:main2} shows a comparison of our results and those expected from five models. In the case of Padova and BaSTI models, we adopted the mean locus line of the massive mock star clusters presented in Figure \ref{fig:stoch}. Note that the Padova02 and BaSTI\_os models show almost same results of AGB luminosity evolution in SSPs. It reflects that the Padova models include the same ingredients as the BaSTI\_os model in terms of the core overshooting. Except for these two models, there are several differences in model expectations. 
First, peak values of the luminosity contribution of AGB stars expected in models are different. The Maraston and Padova10 models suggest 0.7 -- 0.8 for the value of the highest AGB fraction, while other three models (Padova02, BaSTI\_std, and BaSTI\_os) do just up to $\sim$ 0.6. 
Second, all models expect peak values at the similar age range with log(age) $\sim$ 8.6 -- 8.8, except for the BaSTI\_std model. The peak position of the BaSTI model expectation lies at slightly younger age compared with the expectation of other models. It is because the null overshooting efficiency leads to the shorter lifetime of main-sequence stars. 
Third, both young and old parts are different between the Maraston model and other models. Padova02, Padova10, and BaSTI\_os models show that the AGB luminosity fraction in SSPs is lowest at log(age) $\sim$ 8.0 and increases continuously up to the highest value as SSPs are getting older, while the Maraston model suggests that it comes up to already $\sim$ 0.4 at log(age) = 8.0 -- 8.3 and increases drastically afterwards. The BaSTI\_std model, however, expects higher AGB luminosity contribution than any other models at log(age) $\sim$ 8.0 as mentioned above. In the case of SSPs older than 1 Gyr, the AGB luminosity contribution decreases in all models, but the Maraston model shows the steepest decrease.

The bright cluster sequence represents well-populated systems that are less influenced by the stochastic fluctuation than faint star clusters, so that it is more appropriate to compare with the expectation of SSP models. In Figure \ref{fig:main2}, we notice two points with regard to the comparison between the bright cluster sequence and the model expectation: (1) the maximum value of the bright cluster sequence and (2) the age range corresponding to the maximum AGB luminosity contribution.
First, the peak value of the AGB luminosity contribution for bright clusters is up to 0.7 -- 0.8. Only two models, Maraston and Padova10 models, reproduce this maximum value. Second, the peak position for the bright cluster sequence appears at log(age) = 8.5 -- 8.7, which is slightly younger than for model predictions (log(age) = 8.6 -- 8.8). However, this discrepancy is not significant because the errors of ages are around 0.1.

	\subsection{Implication for the study of size and mass evolution of galaxies}
	
The calibration of the EPS models influences galaxy studies because the determination of physical parameters of galaxies depends on the amount of the AGB near-infrared luminosity contribution in EPS models.
The difference between the AGB luminosity fractions expected by Padova02 and Maraston models is largest at log(age) $\sim$ 8.7 -- 8.9 (age $\sim$ 0.5 -- 0.8 Gyr), as shown Figure 13. This indicates that the differences of galaxy mass estimates based on the EPS models can be significant at this age range. This can be important  for galaxies with young stellar ages, such as high-redshift ($z \geq$ 2 -- 3) or post-starburst galaxies. In these kinds of galaxies, the young stellar component with ages $\sim$ 1 Gyr is dominant.
For example, \citet{rai11} presented SED fitting results of 79 early-type galaxies at $z \sim$ 1.3, finding the differences in stellar ages and masses of galaxies estimated with the BC03 and Maraston models, respectively. 
They showed that the masses of galaxies derived with the Maraston model tend to be lower than those estimated with the BC03 model, and this discrepancy is prominent (by a factor of two) at galaxy ages $\sim$ 1.0 -- 1.3 Gyr with uncertainty of $\sim$ 1.0 -- 1.5 Gyr (see Figure 5 in \citealt{rai11}). 

\citet{muz09} reported how the galaxy evolution process depends on EPS models. They investigated the size and mass growth of high-redshift galaxies ($z \sim$ 2.3). 
From the galaxy mass differences based on the EPS models, 
they found that the galaxy size growth from $z \sim$ 2.3 to $z =$ 0 is faster in the case of the BC03 model than the case of the Maraston model, assuming that the mass growth rate is same in these two cases. Thus, the size-mass relation of galaxies can be influenced by the AGB luminosity contribution in each EPS model. 
Our results of the bright cluster sequence are closer to the Maraston model so that they support the size-mass relation of galaxies derived with this model.

\section{Summary and Conclusion}

We investigated the $K_s$ band luminosity evolution of the AGB population in SSPs using 102 LMC star clusters.
First, we determined ages and foreground reddening of star clusters from the $UBV$ photometry in the MCPS \citep{zar04} using Padova isochrones \citep{mar08}.
Then AGB stars in each cluster were selected using 2MASS $(K_s - (J-K_s ))$ CMDs. 
We derived the $K_s$ band luminosity fraction of AGB stars in 102 star clusters as a function of ages.
The $K_s$ band luminosity fraction of AGB stars in star clusters increases as age increases from log(age) $\sim$ 8.0. It reaches a maximum up to $\sim$ 0.6 for mean values and $\sim$ 0.8 for bright cluster sequences at log(age) $\sim$ 8.5, and decreases afterwards. The AGB luminosity fraction for given ages shows a large scatter caused by stochastic effects.
We compared our results with five SSP models: Padova02, Padova10, Maraston, BaSTI\_std, and BaSTI\_os models.
It is found that the only two models (Padova10 and Maraston models) match approximately the observational $K_s$ band AGB luminosity contribution of bright star clusters derived in this study, while other models predict the AGB luminosity contribution much lower.

\acknowledgments
This work was supported by the National Research Foundation of Korea (NRF) grant
funded by the Korea Government (MEST) (No. 2012R1A4A1028713).


\clearpage






\clearpage
	
\begin{deluxetable}{l c c c c c}
\tablecaption{Physical parameters of 102 star clusters in the LMC\label{tab:sc.prop}}
\tablewidth{0pt}
\tablehead{
\colhead{Cluster} & \colhead{$\alpha$ (J2000)} & \colhead{$\delta$ (J2000)} & 
\colhead{Radius [$\arcsec$]} & \colhead{log(age [yr])} & \colhead{$E(B-V)$}
}
\startdata
SL8 & 04:37:51.07 & -69:01:42.9 & 44.0 $\pm$ 4.0 & 8.70 $\pm$ 0.10 & 0.10 \\
NGC 1693 & 04:47:39.92 & -69:20:37.9 & 52.0 $\pm$ 4.0 & 8.00 $\pm$ 0.10 & 0.20 \\
HS37 & 04:50:29.37 & -68:42:45.2 & 30.75$^\emph{a}$ & 8.65 $\pm$ 0.05 & 0.12 \\
SL70 & 04:52:51.01 & -67:23:52.1 & 52.0 $\pm$ 4.0 & 8.70 $^{+ 0.10}_{- 0.05}$ & 0.07 \\
SL117 & 04:56:22.55 & -68:58:01.9 & 85.5 $\pm$ 4.5 & 8.25 $\pm$ 0.05 & 0.12 \\
\enddata
\tablecomments{Table \ref{tab:sc.prop} is published in its entirety in the 
electronic edition.  The five sample star clusters are shown here regarding its form and content.\\
\emph{a.} The radii without errors are adopted from \citet{bic08}.}
\end{deluxetable}

\begin{deluxetable}{l c c c c c c c c}
\rotate
\tabletypesize{\footnotesize}
\tablecaption{ The properties of AGB populations in 102 star clusters in the LMC.
\label{tab:agb.lf}}
\tablewidth{0pt}
\tablehead{
\colhead{Cluster} & 
\colhead{$K_s$} & 
\colhead{$\mathrm{L^{K_s}_{tot}}$ (10$^4$ L$_{\odot}$)} & 
\colhead{N$_{\mathrm{AGB}}$} & 
\colhead{$\mathrm{N_{AGB}^{field}}$} & 
\colhead{$\mathrm{L_{AGB}^{K_s}}$ (10$^4$ L$_{\odot}$)} &
\colhead{$\mathrm{L_{AGB}^{K_s}}$ (10$^4$ L$_{\odot}$) (net)} &
\colhead{$\mathrm{L^{K_s}_{AGB}/L^{K_s}_{tot}}$} &
\colhead{$\mathrm{L^{K_s}_{AGB}/L^{K_s}_{tot}}$ (net)}
}
\startdata
SL8 & 10.429 $\pm$ 0.027 & 3.44 $^{+ 0.55}_{- 0.36}$ & 1 & 0.10 & 1.59 $\pm$ 1.59 & 1.44 $\pm$ 1.51 & 0.462 $^{+ 0.468}_{- 0.464}$ & 0.418 $^{+ 0.444}_{- 0.441}$ \\
NGC 1693 & 8.795 $\pm$ 0.009 & 15.49 $^{+ 0.52}_{- 0.77}$ & 2 & 0.20 & 3.78 $\pm$ 2.67 & 3.40 $\pm$ 2.54 & 0.244 $\pm$ 0.173 & 0.220 $\pm$ 0.164 \\
HS37 & 10.129 $\pm$ 0.015 & 4.53 $^{+ 0.15}_{- 0.08}$ & 1 & 0.14 & 3.49 $\pm$ 3.49 & 2.99 $\pm$ 3.23 & 0.771 $\pm$ 0.771 & 0.660 $^{+ 0.714}_{- 0.713}$ \\
SL70 & 9.858 $\pm$ 0.019 & 5.82 $^{+ 1.99}_{- 1.73}$ & 2 & 0.39 & 4.09 $\pm$ 2.89 & 3.30 $\pm$ 2.60 & 0.703 $^{+ 0.552}_{- 0.539}$ & 0.567 $^{+ 0.487}_{- 0.477}$ \\
SL117 & 8.359 $\pm$ 0.011 & 23.14 $^{+ 2.05}_{- 3.37}$ & 4 & 2.01 & 15.6 $\pm$ 7.8 & 7.77 $\pm$ 5.50 & 0.674 $^{+ 0.342}_{- 0.351}$ & 0.336 $^{+ 0.240}_{- 0.243}$ \\
\enddata
\tablecomments{Table \ref{tab:agb.lf} is published in its entirety in the 
electronic edition.  
The five sample star clusters are shown here regarding its form and content.
$\mathrm{L_{AGB}^{K_s}}$ (net) and $\mathrm{L^{K_s}_{AGB}/L^{K_s}_{tot}}$ (net) indicate the field-subtracted values.}

\end{deluxetable}


\begin{deluxetable}{c c c c c c c c c c c}
\rotate
\tabletypesize{\scriptsize}
\tablecaption{Comparison of $K_s$ band luminosity fraction of AGB stars with SSP models.
\label{tab:comp.lf}}
\tablewidth{0pt}
\tablehead{
\multicolumn{1}{c}{ } & \multicolumn{4}{c}{Observation (This study)} & \colhead{} & \multicolumn{5}{c}{Model expectation} \\
\cline{2-5} \cline{7-11}\\
\colhead{ } & \colhead{} & & \colhead{Bright cluster} & \colhead{Bright cluster} & \colhead{} & \colhead{Padova} & \colhead{Padova} & \colhead{Maraston} & \colhead{BaSTI\_std} & \colhead{BaSTI\_os} \\

\colhead{log(age [yr])} & \colhead{Mean}  & \colhead{Mean (net)} & \colhead{sequence} & \colhead{sequence (net)} & \colhead{} & \colhead{02} & \colhead{10} & \colhead{} & \colhead{(standard)} & \colhead{(overshooting)}

}
\startdata
8.00 & 0.25$^{+ 0.06}_{- 0.07}$ & 0.24 $\pm$ 0.07 & 0.30 & 0.20 & & 0.21 & 0.17 & 0.40 & 0.43 & 0.26 \\
8.25 & 0.53 $\pm$ 0.11 & 0.38 $\pm$ 0.10 & 0.62 & 0.50 & & 0.43 & 0.52 & 0.41 & 0.58 & 0.50 \\
8.50 & 0.54 $\pm$ 0.11 & 0.47 $\pm$ 0.11 & 0.81 & 0.67 & & 0.51 & 0.62 & 0.73 & 0.62 & 0.55 \\
8.75 & 0.57$^{+ 0.14}_{- 0.13}$ & 0.52 $\pm$ 0.13 & 0.78 & 0.70 & & 0.55 & 0.71 & 0.78 & 0.57 & 0.58 \\
9.00 & 0.52 $\pm$ 0.11 & 0.38 $\pm$ 0.09 & 0.60 & 0.55 & & 0.51 & 0.66 & 0.62 & 0.50 & 0.53 \\
9.25 & 0.31 $\pm$ 0.09 & 0.27 $\pm$ 0.08 & 0.33 & 0.30 & & 0.38 & 0.45 & 0.28 & 0.34 & 0.40 \\
9.50 & 0.32 $\pm$ 0.19 & 0.21 $\pm$ 0.15 & 0.20 & 0.15 & & 0.29 & 0.36 & 0.12 & 0.30 & 0.32 \\
\enddata

\tablecomments{Mean (net) and bright cluster sequence (net) indicate the field-subtracted values.}

\end{deluxetable}

\end{document}